\documentclass[aps,prl,twocolumn,superscriptaddress,showpacs,preprintnumbers,amsmath,amssymb]{revtex4}

\usepackage{graphicx}
\usepackage{dcolumn}
\graphicspath{{ps/}}

\renewcommand{\arraystretch}{1.1}

\begin{document}

\title{\quad\\[1.0cm] Study of the {\boldmath $B\to X(3872)(\to D^{*0}\bar D^0)K$} decay}

\affiliation{Budker Institute of Nuclear Physics, Novosibirsk}
\affiliation{University of Cincinnati, Cincinnati, Ohio 45221}
\affiliation{Justus-Liebig-Universit\"at Gie\ss{}en, Gie\ss{}en}
\affiliation{The Graduate University for Advanced Studies, Hayama}
\affiliation{Gyeongsang National University, Chinju}
\affiliation{Hanyang University, Seoul}
\affiliation{University of Hawaii, Honolulu, Hawaii 96822}
\affiliation{High Energy Accelerator Research Organization (KEK), Tsukuba}
\affiliation{Institute of High Energy Physics, Chinese Academy of Sciences, Beijing}
\affiliation{Institute of High Energy Physics, Vienna}
\affiliation{Institute of High Energy Physics, Protvino}
\affiliation{Institute for Theoretical and Experimental Physics, Moscow}
\affiliation{J. Stefan Institute, Ljubljana}
\affiliation{Kanagawa University, Yokohama}
\affiliation{Institut f\"ur Experimentelle Kernphysik, Karlsruhe Institut f\"ur Technologie, Karlsruhe}
\affiliation{Korea University, Seoul}
\affiliation{Kyungpook National University, Taegu}
\affiliation{\'Ecole Polytechnique F\'ed\'erale de Lausanne (EPFL), Lausanne}
\affiliation{Faculty of Mathematics and Physics, University of Ljubljana, Ljubljana}
\affiliation{University of Maribor, Maribor}
\affiliation{Max-Planck-Institut f\"ur Physik, M\"unchen}
\affiliation{University of Melbourne, School of Physics, Victoria 3010}
\affiliation{Nagoya University, Nagoya}
\affiliation{Nara Women's University, Nara}
\affiliation{National Central University, Chung-li}
\affiliation{National United University, Miao Li}
\affiliation{Department of Physics, National Taiwan University, Taipei}
\affiliation{H. Niewodniczanski Institute of Nuclear Physics, Krakow}
\affiliation{Nippon Dental University, Niigata}
\affiliation{Niigata University, Niigata}
\affiliation{Novosibirsk State University, Novosibirsk}
\affiliation{Osaka City University, Osaka}
\affiliation{Panjab University, Chandigarh}
\affiliation{University of Science and Technology of China, Hefei}
\affiliation{Seoul National University, Seoul}
\affiliation{Sungkyunkwan University, Suwon}
\affiliation{School of Physics, University of Sydney, NSW 2006}
\affiliation{Tata Institute of Fundamental Research, Mumbai}
\affiliation{Excellence Cluster Universe, Technische Universit\"at M\"unchen, Garching}
\affiliation{Toho University, Funabashi}
\affiliation{Tohoku Gakuin University, Tagajo}
\affiliation{Department of Physics, University of Tokyo, Tokyo}
\affiliation{Tokyo Metropolitan University, Tokyo}
\affiliation{Tokyo University of Agriculture and Technology, Tokyo}
\affiliation{IPNAS, Virginia Polytechnic Institute and State University, Blacksburg, Virginia 24061}
\affiliation{Yonsei University, Seoul}
  \author{T.~Aushev}\affiliation{\'Ecole Polytechnique F\'ed\'erale de Lausanne (EPFL), Lausanne}\affiliation{Institute for Theoretical and Experimental Physics, Moscow} 
  \author{N.~Zwahlen}\affiliation{\'Ecole Polytechnique F\'ed\'erale de Lausanne (EPFL), Lausanne} 
  \author{I.~Adachi}\affiliation{High Energy Accelerator Research Organization (KEK), Tsukuba} 
  \author{H.~Aihara}\affiliation{Department of Physics, University of Tokyo, Tokyo} 
  \author{A.~M.~Bakich}\affiliation{School of Physics, University of Sydney, NSW 2006} 
  \author{V.~Balagura}\affiliation{Institute for Theoretical and Experimental Physics, Moscow} 
  \author{A.~Bay}\affiliation{\'Ecole Polytechnique F\'ed\'erale de Lausanne (EPFL), Lausanne} 
  \author{K.~Belous}\affiliation{Institute of High Energy Physics, Protvino} 
  \author{V.~Bhardwaj}\affiliation{Panjab University, Chandigarh} 
  \author{M.~Bischofberger}\affiliation{Nara Women's University, Nara} 
% ...
\collaboration{The Belle Collaboration}


\begin{abstract}
We present a study of $B\to X(3872)K$ with $X(3872)$ decaying to
$D^{*0}\bar D^0$ using a sample of 657 million $B\bar B$ pairs
recorded at the $\Upsilon(4S)$ resonance with the Belle detector at
the KEKB asymmetric-energy $e^+e^-$ collider.  Both $D^{*0}\to
D^0\gamma$ and $D^{*0}\to D^0\pi^0$ decay modes are used.  We find a
peak of $50.1^{+14.8}_{-11.1}$ events with a mass of
$(3872.9^{+0.6\,+0.4}_{-0.4\,-0.5}){\rm~MeV}/c^2$, a width of
$(3.9^{+2.8\,+0.2}_{-1.4\,-1.1}){\rm~MeV}/c^2$ and a product branching
fraction ${\cal B}(B\to X(3872)K)\times{\cal B}(X(3872)\to D^{*0}\bar
D^0)=(0.80\pm0.20\pm0.10)\times10^{-4}$, where the first errors are
statistical and the second ones are systematic.  The significance of
the signal is $6.4\sigma$.  The difference between the fitted mass and
the $D^{*0}\bar D^0$ threshold is calculated to be
$(1.1^{+0.6\,+0.1}_{-0.4\,-0.3}){\rm~MeV}/c^2$.  We also obtain an
upper limit on the product of branching fractions ${\cal B}(B\to
Y(3940)K)\times{\cal B}(Y(3940)\to D^{*0}\bar D^0)$ of
$0.67\times10^{-4}$ at 90\%~CL.
\end{abstract}

\maketitle

The $X(3872)$ was discovered by Belle in 2003 in $B^\pm\to
J/\psi\pi^+\pi^-K^\pm$~\cite{x-discovery} with a mass of
$(3872.0\pm0.6\pm0.5){\rm~MeV}/c^2$, and was later confirmed by
CDF~\cite{x-CDF}, D\O~\cite{x-D0} and BaBar~\cite{x-babar-confirm}.
It is one of the many new and unexpected hidden-charm states states
recently discovered with masses around $4{\rm~GeV}/c^2$.  So far it
remains unclassified; it does not seem to be a pure $c\bar c$
charmonium state, but may be a $D^*\bar D$
deuson~\cite{x-swanson,x-tornqvist}, a tetraquark
state~\cite{x-maiani} or a charmonium-gluon hybrid~\cite{qqg-li}.  The
current average mass in the $J/\psi\pi^+\pi^-$ channel is
$(3871.50\pm0.19){\rm~MeV}/c^2$~\cite{mXaverage}.

An important feature of the $X(3872)$ is that its mass is very close
to the $D^{*0}\bar D^0$ threshold
($(3871.81\pm0.36){\rm~MeV}/c^2$~\cite{PDG08}).  The $X(3872)$ was
also observed by Belle~\cite{x-gokhroo} as a near-threshold
enhancement in the $D^0\bar D^0\pi^0$ invariant mass spectrum of the
$B\to D^0\bar D^0\pi^0K$ channel, with a peak at
$(3875.2\pm0.7^{+0.3}_{-1.6}\pm0.8){\rm~MeV}/c^2$, where the third
error is from the uncertainty on the $D^0$ mass~\cite{PDG08}, a
Gaussian width of $(2.42\pm0.55){\rm~MeV}/c^2$, and a branching
fraction ${\cal B}(B\to D^0\bar
D^0\pi^0K)=(1.22\pm0.31^{+0.23}_{-0.30})\times10^{-4}$.  This initial
study did not distinguish between decays via the $D^{*0}$, and more
general $D^0\pi^0$ final states.  Looking in addition for the
$D^{*0}\to D^0\gamma$ decay is crucial to demonstrate the presence of
$X(3872)$ decay through a $D^{*0}$.

The BaBar collaboration recently published an observation of the decay
$B\to X(3872)(\to D^{*0}\bar D^0)K$ with a $4.9\sigma$
significance~\cite{x-babar}.  The observed mass is
$(3875.1^{+0.7}_{-0.5}\pm0.5){\rm~MeV}/c^2$ and the width is
$(3.0^{+1.9}_{-1.4}\pm0.9){\rm~MeV}/c^2$, with a product branching
fraction ${\cal B}(B^+\to X(3872)K^+)\times{\cal B}(X(3872)\to
D^{*0}\bar D^0)=(1.67\pm0.36\pm0.47)\times10^{-4}$.  In the BaBar
analysis, $D^{*0}$ candidates were subjected to a mass-constrained
fit.

Another new particle called $X(3940)$ was discovered by Belle in the
$e^+e^-\to D\bar D^{*0}, J/\psi$ process~\cite{x3940}.  A state with
the same mass, the $Y(3940)$ (also known as $X(3945)$~\cite{PDG08}), was 
discovered by Belle in $B\to\omega J/\psi K$~\cite{Y3940-belle} and
was later confirmed by BaBar, albeit with a smaller
mass~\cite{Y3940-babar}.  The possibility that the $X(3940)$ and 
$Y(3940)$ are the same state has not yet been ruled out.

In this paper we measure the $X(3872)\to D^{*0}\bar D^0$ decay mode,
followed either by $D^{*0}\to D^0\gamma$ or $D^{*0}\to D^0\pi^0$, in
charged and neutral $B\to X(3872)K$ decays.  Inclusion of
charge-conjugate modes is implied throughout the paper.  Furthermore,
we use the notation $D^{*0}\bar D^0$ to indicate both $D^{*0}\bar D^0$
and $\bar D^{*0}D^0$.  The results are based on a $605{\rm~fb}^{-1}$
data sample, corresponding to $657\times10^6\,B\bar B$ pairs,
collected at the $\Upsilon(4S)$ resonance with the Belle
detector~\cite{BelleDet} at the KEKB asymmetric-energy $e^+e^-$
collider~\cite{KEKB}, which includes the statistics used in the
previous Belle analysis~\cite{x-gokhroo}.

The Belle detector is a general purpose spectrometer with a
$1.5{\rm~T}$ magnetic field provided by a superconducting solenoid.  A
silicon vertex detector and a 50-layer central drift chamber are used
to measure the momenta of charged particles.  Photons are detected in
an electromagnetic calorimeter consisting of CsI(Tl) crystals.
Particle identification likelihoods ${\cal L}_K$ and ${\cal L}_{\pi}$
are derived from information provided by an array of time-of-flight
counters, an array of silica aerogel Cherenkov threshold counters and
${\rm d}E/{\rm d}x$ measurements in the central drift chamber.

Charged tracks are identified as kaons using a requirement on the
likelihood ratio ${\cal L}_K/({\cal L}_K+{\cal L}_{\pi})$, which has
an efficiency of $88\%$ for kaons and $10\%$ for pions.  Similarly,
the charged pion selection has an efficiency of $98\%$ for pions and
$12\%$ for kaons.  $\pi^0$ candidates are reconstructed from pairs of
photons with energies $E_{\gamma}>50{\rm~MeV}$ and with invariant mass
in the range $118{\rm~MeV}/c^2<M_{\gamma\gamma}<150{\rm~MeV}/c^2$.  A
mass-constrained fit is applied to obtain the four-momenta of the
$\pi^0$ candidates.  $K_S^0$ candidates are reconstructed in the
$K_S^0\to\pi^+\pi^-$ mode with the requirement
$|M_{\pi\pi}-m_{K_S^0}|<15{\rm~MeV}/c^2$, where $m_{K_S^0}$ is the
$K_S^0$ mass~\cite{PDG08}.  Requirements on the $K_S^0$ vertex
displacement from the interaction point and on the difference between
vertex and $K_S^0$ flight directions are applied.  The $K_S^0$
selection criteria are described in detail elsewhere~\cite{FangKs}.  A
mass- and vertex-constrained fit is applied to improve the
four-momenta measurements of the $K_S^0$ candidates.

$D^0$ mesons are reconstructed in the modes $D^0\to K^-\pi^+$,
$K^-\pi^+\pi^0$, $K^-\pi^+\pi^+\pi^-$, $K_S^0\pi^+\pi^-$ and $K^-K^+$.
The $D^0$ candidates are required to have a mass within
$\pm14{\rm~MeV}/c^2$ ($\pm26{\rm~MeV}/c^2$ for the $K^-\pi^+\pi^0$
mode) of the $D^0$ mass, $1864.8{\rm~MeV}/c^2$~\cite{PDG08}.  This
mass window width corresponds to $\pm4\sigma$ ($\pm3\sigma$ for
$K^-\pi^+\pi^0$).  Mass- and vertex-constrained fits are applied to
improve the resolution of the four-momenta of $D^0$ candidates.  $D^0$
candidates are combined with a photon or a $\pi^0$ to obtain $D^{*0}$
candidates.  The photon candidate is required to have an energy in
excess of $100{\rm~MeV}$ and shower shape variables that are
consistent with an electromagnetic shower; the ratio of energy
deposition in the central $3\times3$ and $5\times5$ crystals of the
cluster is required to be greater than 0.8.  A mass window of
$\pm27.5{\rm~MeV}/c^2$ for the $D^{*0}\to D^0\gamma$ channel and
$\pm6{\rm~MeV}/c^2$ for the $D^{*0}\to D^0\pi^0$ channel is applied
($\pm3\sigma$), and a mass-constrained fit is used to improve the
four-momenta of the $D^{*0}$ candidates; the mass is constrained to
$2007.0{\rm~MeV}/c^2$~\cite{PDG08}.

$B$ mesons are reconstructed by combining a $D^{*0}\bar D^0$ candidate
with invariant mass $M_{D^*D}<4.0{\rm~GeV}/c^2$ and a charged or
neutral kaon candidate.  To further reduce the background, at least
one $D^0$($\bar D^0$) is required to decay to $K^-\pi^+$($K^+\pi^-$).
The beam-energy constrained mass $M_{\rm bc}=\sqrt{E_{\rm
    beam}^2-\left(\sum_i\vec{P}_i\right)^2}$, where $\vec{P}_i$ is the
momentum of the $i$th daughter of the $B$ candidate in the $e^+e^-$
center-of-mass (CM) system, is required to be larger than
$5.2{\rm~GeV}/c^2$.  The energy difference $\Delta E=E_B-E_{\rm
  beam}$, where $E_B$ is the CM energy of the $B$ candidate and
$E_{\rm beam}$ is the CM beam energy, is restricted to the range
$|\Delta E|<25{\rm~MeV}/c^2$.  Continuum $e^+e^-\to q\bar q$
background events ($q=u,d,s,c)$ are suppressed by requiring the ratio
of the second and zeroth Fox-Wolfram moments~\cite{fox} to be smaller
than 0.3.

The average $B$ candidate multiplicity per event is $2.3$ for
$D^{*0}\to D^0\gamma$ and $2.7$ for $D^{*0}\to D^0\pi^0$.  We select
the candidate with the smallest value of the quantity
\begin{eqnarray}
  \left(\frac{\Delta M_{D^0}}{\sigma_{M_{D^0}}}\right)^2 +
  \left(\frac{\Delta M_{\bar D^0}}{\sigma_{M_{D^0}}}\right)^2 +
  \left(\frac{\Delta(M_{D^{*0}}-M_{D^0})}{\sigma_{(M_{D^{*0}}-M_{D^0})}}\right)^2
\nonumber\\
  + \left(\frac{\Delta E}{\sigma_{\Delta E}}\right)^2 \; 
  \left[+\left(\frac{\Delta M_{\pi^0}}{\sigma_{M_{\pi^0}}}\right)^2\right],\;
\end{eqnarray}
where $\Delta x$ is the deviation of the measured quantity $x$ from
its expected value and $\sigma_x$ the uncertainty in its measurement
obtained using a Monte Carlo (MC) method, and the last term applies to
the $D^{*0}\to D^0\pi^0$ channel only.

MC samples are produced using the EVTGEN~\cite{evtgen} generator.  The
$X(3872)$ mass distribution is generated according to a relativistic
Breit-Wigner function
\begin{eqnarray}
\label{eqn:relBW}
BW(m) & = & \frac{\mu m\Gamma(m)}{(m^2-\mu^2)^2+\mu^2\Gamma(m)^2}\,,\hspace{1.5cm}\\
\rm{where}\hspace{.7cm}
\quad\Gamma(m) & = & \Gamma_0\frac{\mu}{m}\frac{p(m)}{p(\mu)}, 
\nonumber\\
p(m)  & = & \frac{1}{2m}\sqrt{\left(m^2-(m_{D^0}+m_{D^{*0}})^2 \right)}
\nonumber\\
      &   & \times\sqrt{\left(m^2-(m_{D^0}-m_{D^{*0}})^2 \right)},
\nonumber
\end{eqnarray}
and $\mu$ and $\Gamma_0$ are the nominal mass and width of the
resonance, respectively, and $p(m)$ is the momentum of one of the
daughters in the rest frame of its parent.  The term $m\Gamma(m)$ in
the numerator of Eq.~(\ref{eqn:relBW}) behaves like a phase-space
function, giving a smooth rise near the $D^{*0}\bar D^0$ threshold.

\begin{figure}[htbp]
  \begin{center}
    \includegraphics[width=0.49\textwidth]{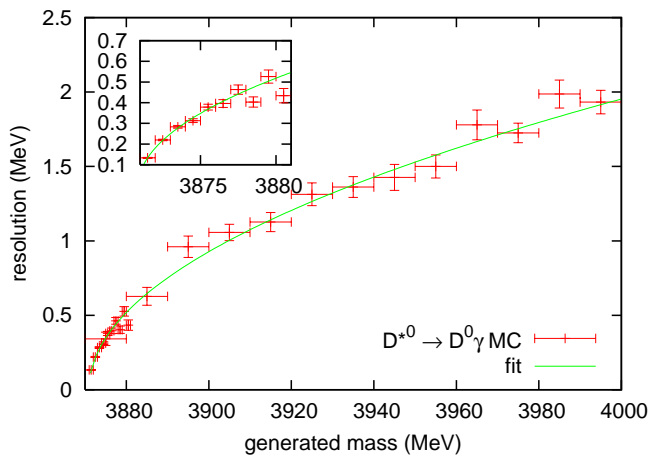}
    \caption{$X(3872)$ mass resolution as a function of the $X(3872)$
      mass in the $D^0\gamma$ channel, obtained from MC with a
      $X(3872)$ mass spectrum generated for a continuous range of
      masses from threshold to $4.0{\rm~GeV}/c^2$.  Crosses are
      Gaussian resolutions for various generated $D^{*0}\bar D^0$
      masses; the curve is the result of a fit with a square root
      function.  Very similar results are obtained for the $D^0\pi^0$
      channel.}
    \label{fig:resol}
  \end{center}
\end{figure}

The response of the Belle detector is simulated using a GEANT3-based
program~\cite{geant3}.  Since the $X(3872)$ is very close to the
$D^{*0}\bar D^0$ threshold, its mass resolution varies rapidly with
the $D^{*0}\bar D^0$ mass ($M_{D^*\bar D}$).  This resolution was
studied in detail using MC simulations.  It is modeled as a single
Gaussian with a width given by the function $a\sqrt{M_{D^*D}-m_0}$
(where $a$ and $m_0$ are free parameters) shown in
Fig.~\ref{fig:resol}.  At $3872{\rm~MeV}/c^2$, the resolution is
$0.13{\rm~MeV}/c^2$.

A two-dimensional unbinned extended maximum likelihood fit to $M_{\rm
  bc}$ and $M_{D^*D}$ is performed.  The $M_{\rm bc}$ distribution is
described by a single Gaussian function for the signal and an ARGUS
function~\cite{argus} for the background; the $M_{D^*D}$ distribution
is described by a relativistic Breit-Wigner function convoluted with
the mass-dependent Gaussian resolution for the signal and a square
root function for the background.  In the $D^0\gamma$ channel, the
signal function also includes a broad higher mass distribution,
corresponding to $D^{*0}(D^0\gamma)\bar D^0$ events incorrectly
reconstructed as $D^0\bar D^{*0}(\gamma\bar D^0)$.  The shape and
fraction of this contribution is determined from MC.  In the
$D^0\pi^0$ channel, however, the reflection shape is too similar to
the signal one to be distinguished.  This is taken into account as a
systematic uncertainty.  Additional components are the $Y(3940)$
signal, described by a relativistic Breit-Wigner function, and the
non-resonant $B\to D^*\bar DK$ background, which peaks in $M_{\rm bc}$
but not in $M_{D^*D}$ and is therefore described by a single Gaussian
function in $M_{\rm bc}$ and a square root function in $M_{D^*D}$.
The fitting procedure has been extensively tested using toy MC
samples.

First, each $D^{*0}$ decay channel sample is fitted separately.  The
yield, mass and width of the $X(3872)$ peak are free parameters of the
fit, as well as the $Y(3940)$ yield and the number of background and
non-resonant $B\to D^*\bar DK$ events.  The $Y(3940)$ mass and its
width are fixed to the values of Ref.~\cite{Y3940-belle}.  The results
of the fits are presented in Table~\ref{tab:results}.  Since all the
results are consistent we proceed with a combined fit.

\begin{table*}[htb]
  \begin{center}
    \renewcommand{\arraystretch}{1.4}
    \caption{Summary of results: the fitted mass, width, and yield of
      the $X(3872)$ peak, and the total reconstruction efficiency,
      branching fraction and statistical significance for the various
      fits described in the text.}
    \label{tab:results}
    \begin{ruledtabular}
      \begin{tabular}{ccccccc}
        Sample & $M_X$ (${\rm MeV}/c^2$) & $\Gamma$ (${\rm MeV}/c^2$) 
        & Yield & $\epsilon\times{\cal B}$ & ${\cal B}$ ($10^{-4}$) & $\sigma$ \\
        \hline
        $D^{*0}\to D^0\gamma$ ($XK^+$ and $XK^0$) & $3873.4\pm1.0$ & $4.2^{+3.7}_{-1.8}$ 
        & $26.2^{+9.0}_{-7.6}$ & $4.56\times10^{-4}$ & $0.87\pm0.28\pm0.10$ & $4.4\sigma$\\
        $D^{*0}\to D^0\pi^0$ ($XK^+$ and $XK^0$) & $3872.8\pm0.7$ & $3.1^{+4.1}_{-1.5}$
        & $22.0^{+10.7}_{-6.4}$ & $4.93\times10^{-4}$ & $0.68\pm0.26\pm0.09$ & $6.8\sigma$\\
        \hline
        All (free $D^0\gamma/D^0\pi^0$ ratio) & $3872.9^{+0.6}_{-0.4}$ & $3.9^{+2.7}_{-1.4}$ 
        & $50.6^{+14.2}_{-11.0}$ & $9.49\times10^{-4}$ & $0.81\pm0.20\pm0.10$ & $7.9\sigma$\\
        All (fixed $D^0\gamma/D^0\pi^0$ ratio)& $3872.9^{+0.6}_{-0.4}$ & $3.9^{+2.8}_{-1.4}$ 
        & $50.1^{+14.8}_{-11.1}$ & $9.49\times10^{-4}$ & $0.80\pm0.20\pm0.10$ & $7.9\sigma$\\
        \hline
        $B^+\to XK^+$ & $3872.9$ (fixed) & $3.9$ (fixed) & $41.3^{+9.1}_{-8.1}$ 
        & $8.17\times10^{-4}$ & $0.77\pm0.16\pm0.10$ & $7.6\sigma$ \\
        $B^0\to XK^0$ & $3872.9$ (fixed) & $3.9$ (fixed) &  $8.4^{+4.5}_{-3.6}$ 
        & $1.32\times10^{-4}$ & $0.97\pm0.46\pm0.13$ & $2.8\sigma$ \\
      \end{tabular}
      \renewcommand{\arraystretch}{1.0}
    \end{ruledtabular}
  \end{center}
\end{table*}

\begin{figure*}[htbp]
  \begin{center}
    \includegraphics[width=\textwidth]{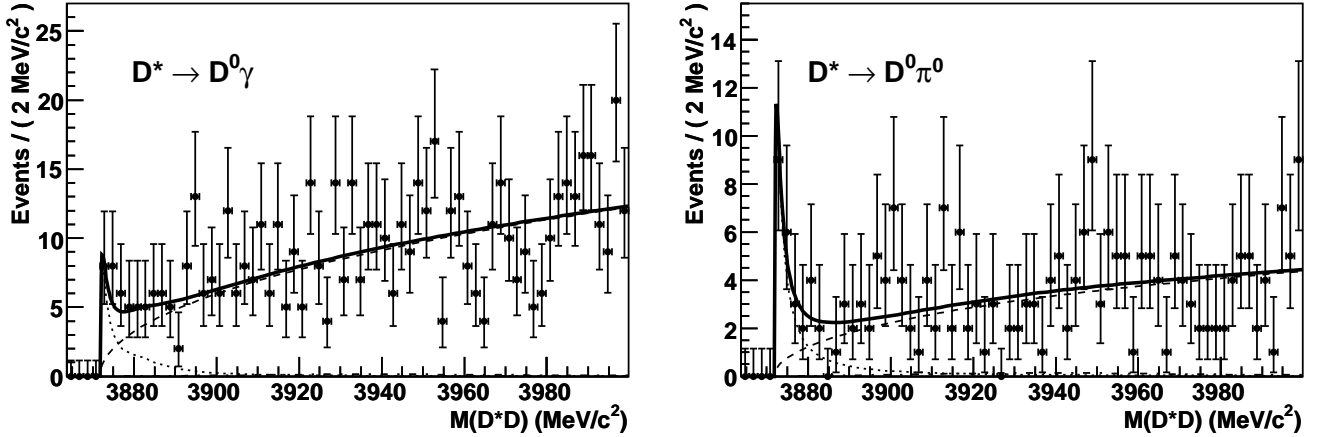}
    \vspace{-7mm}
    \caption{Distributions of $M_{D^*D}$ mass for $M_{\rm
        bc}>5.27{\rm~GeV}/c^2$ for $D^{*0}\to D^0\gamma$ (left) and
      for $D^{*0}\to D^0\pi^0$ (right).  The result of the
      simultaneous fit is shown by the superimposed lines.  The points
      with error bars are data, the dotted curve is the signal, the
      dashed curve is the background, the dash-dotted curve (barely
      visible) is the $Y(3940)$ component, and the solid curve is the
      total fitting function.}
    \label{fig:signal_mx}
  \end{center}
\end{figure*}

\begin{figure*}[htbp]
  \begin{center}
    \includegraphics[width=\textwidth]{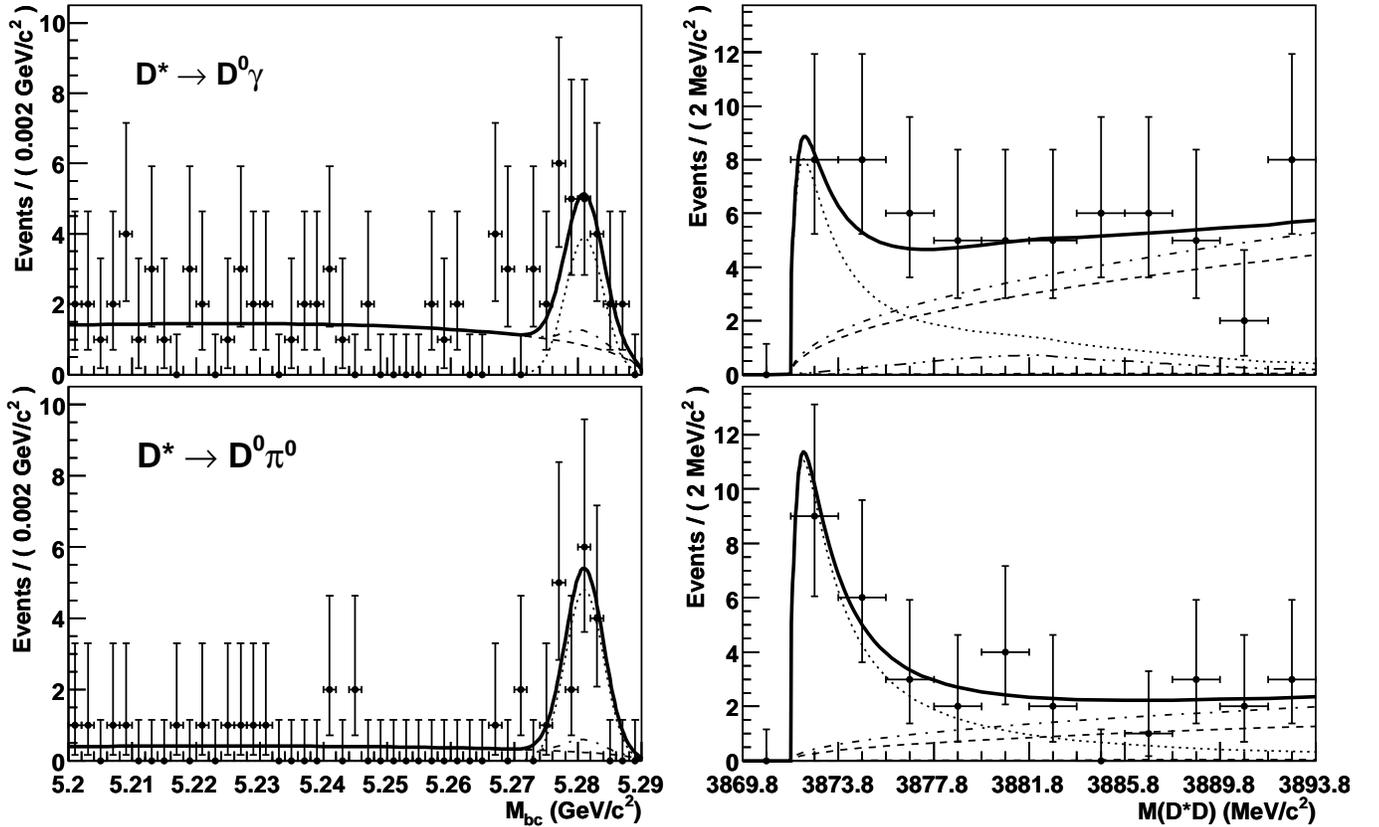}
    \vspace{-7mm}
    \caption{Distributions of $M_{\rm bc}$ for
      $M_{D^*D}<3.88{\rm~GeV}/c^2$ (left) and of $M_{D^*D}$ for
      $M_{\rm bc}>5.27{\rm~GeV}/c^2$ (right); the top row is for
      $D^{*0}\to D^0\gamma$ and the bottom row for $D^{*0}\to
      D^0\pi^0$.  The result of the simultaneous fit is shown by the
      superimposed lines.  The points with error bars are data, the
      dotted curve is the signal, the dashed curve is the background,
      the dash-dotted curve is the sum of the background and the $B\to
      D^*DK$ component, the dot-dot-dashed curve is the contribution
      from $D^0$--$\bar D^0$ reflections, and the solid curve is the
      total fitting function.}
    \label{fig:signal}
  \end{center}
\end{figure*}

We subsequently perform a simultaneous fit to both $D^{*0}$ modes
where the mass and width of the signal function are constrained to
have the same values in both modes, but the ratio of the yields in the
$D^0\gamma$ and $D^0\pi^0$ channels is left free.
Table~\ref{tab:results} shows the fit results.  We obtain a yield
ratio of $N_{D^0\gamma}/N_{D^0\pi^0}=1.16^{+0.53}_{-0.37}$, which is
consistent with the value of $0.92$ we expect from MC with no
non-resonant $D^0\bar D^0\pi^0$ contribution.  We then fix the ratios
of $D^0\gamma/D^0\pi^0$ signal yields for the $X(3872)$ and $Y(3940)$
from MC studies assuming that the $D^0\pi^0$ and $D^0\gamma$ come from
a $D^{*0}$.  This fit gives $50.1^{+14.8}_{-11.1}$ events with a
statistical significance of $7.9\sigma$ (see Fig.~\ref{fig:signal_mx}
and Table~\ref{tab:results}).  We compute the significance from
$-2\ln({\cal L}_0/{\cal L}_{\rm max})$, where ${\cal L}_0$ and ${\cal
  L}_{\rm max}$ are the likelihood values returned by the fit with the
signal yield fixed at zero and left free, respectively.  This quantity
should be distributed as $\chi^2(n_{\rm dof}=3)$, as three parameters
are free for the signal.  The results of the simultaneous fits are
consistent with the results of the individual fits.  The distributions
for $M_{\rm bc}$ and $X(3872)$ mass closer to the $X(3872)$ signal
region are presented in Fig.~\ref{fig:signal}.

Next we fit the $B^+\to X(3872)K^+$ and $B^0\to X(3872)K_S^0$ modes
separately, fixing the $X(3872)$ mass and width to the values obtained
with the simultaneous fit.  Table~\ref{tab:results} shows the results
of these fits.  Assuming the $B^0\to X(3872)K^0$ transition rate to be
equal to twice the $B^0\to X(3872)K_S^0$ rate, we obtain a ratio of
branching fractions
\begin{displaymath}
  \frac{{\cal B}(B^0\to X(3872)K^0)}{{\cal B}(B^+\to X(3872)K^+)} =
  1.26\pm0.65\pm0.06,
\end{displaymath}
which is consistent with unity.  In this ratio, most of the systematic
uncertainties cancel out.  The remaining uncertainties are MC
statistics, particle identification and reconstruction efficiencies of
the $K^+$ and $K_S^0$, which combine in quadrature to a 5\%
uncertainty.

The systematic uncertainties for the mass, width and signal yield are
estimated by varying the fixed parameters of the simultaneous fit: the
resolution function is scaled by factors 0.9 and 1.2 obtained from the
inclusive $D^{*0}$ mass resolution study; the ratio between
$D^0\gamma$ and $D^0\pi^0$ yields is varied according to the error in
the relative branching ratio between $D^{*0}\to D^0\gamma$ and
$D^{*0}\to D^0\pi^0$~\cite{PDG08} and the uncertainty in their
relative reconstruction efficiency; the $Y(3940)$ mass and width are
fixed instead to the values from Ref.~\cite{Y3940-babar}; the yield of
$Y(3940)$ is fixed to zero and to the upper limit of 40 events; the
shape of the combinatorial background is changed from a square root to
an inverted ARGUS function.  Additional uncertainties due to possible
biases of the fit parameters are obtained using a large ensemble of
toy MC experiments generated for different values of $X(3872)$ mass
and width.  We also consider systematic uncertainties due to the
contribution from $D^{*0}\bar D^0\to D^0\bar D^{*0}$
mis-reconstruction by changing the shape of the miss-reconstruction
component to one obtained from MC generated with different values of
$X(3872)$ mass and width.  Using MC we verify that the crossfeeds
between $D^0\bar D^0\gamma$ and $D^0\bar D^0\pi^0$ do not produce
peaks in the signal region.  Another systematic uncertainty on the
mass is due to the $D^{*0}\bar D^0$ threshold mass,
$2m_{D^0}+\Delta(m_{D^{*0}}-m_{D^0})=(3871.80\pm0.41){\rm~MeV}/c^2$~\cite{PDG08}.
The systematic uncertainties for the branching fraction are estimated
from the following sources: the number of $B\bar B$ events, $D^0$
branching fractions, track finding efficiencies, $K/\pi$
identification efficiency, $\gamma$ or $\pi^0$ detection efficiency,
$K_S^0$ reconstruction efficiency, limited MC statistics and variation
of the fixed parameters of the fit.  Table~\ref{tab:syst} summarizes
the systematic uncertainties.  We obtain total systematic errors of
$^{+0.41}_{-0.54}{\rm~MeV}/c^2$ for the mass,
$^{+0.22}_{-1.07}{\rm~MeV}/c^2$ for the width and $\pm12\%$ for the
branching fraction measurement.  The significance of the signal
including systematic uncertainties decreases to $6.4\sigma$.

\begin{table}[htbp]
  \begin{center}
    \caption{Sources of systematic uncertainties for the fitted mass
      and width (${\rm MeV}/c^2$) and for the $B\to X(3872)K$
      branching fraction (\%).}
    \label{tab:syst}
    \begin{ruledtabular}
    \begin{tabular}{lccc}
      Source                             & Mass & Width & ${\cal B}$\\
      \hline
      Resolution function                & $\pm0.04$ & $\pm0.09$ & $\pm0.3$ \\
      $D^0\gamma/D^0\pi^0$ yields ratio  & $\pm0.01$ & $\pm0.00$ & $\pm0.2$ \\
      $Y(3940)$ parameters               & $\pm0.01$ & $^{+0.00}_{-0.32}$ & $^{+0.0}_{-4.0}$ \\
      Background shape                   & $\pm0.00$ & $^{+0.00}_{-0.14}$ & $^{+0.0}_{-2.2}$ \\
      Fit bias                  & $^{+0.05}_{-0.30}$ & $^{+0.15}_{-1.00}$ & $^{+5.0}_{-0.0}$ \\
      $D^0$--$\bar D^0$ reflections      & $\pm0.02$ & $\pm0.11$ & $\pm0.5$ \\
      $D^0$ and $D^{*0}$ masses          & $\pm0.41$ & $-$ & $-$ \\
      Number of $B\bar B$ events         & $-$ & $-$ & $\pm1.4$ \\
      $D^0$ branching fractions          & $-$ & $-$ & $\pm2.4$ \\
      Tracking efficiency                & $-$ & $-$ & $\pm5.0$ \\
      Particle identification            & $-$ & $-$ & $\pm4.0$ \\
      $\gamma$ or $\pi^0$ reconstruction & $-$ & $-$ & $\pm7.3$ \\
      $K_S^0$ reconstruction             & $-$ & $-$ & $\pm4.5$ \\
      MC statistics (efficiency)         & $-$ & $-$ & $\pm1.4$ \\
      \hline
      Total (quadrature)                 & $^{+0.41}_{-0.54}$ & $^{+0.22}_{-1.07}$ & $\pm12$ \\
    \end{tabular}
    \end{ruledtabular}
  \end{center}
\end{table}

In order to determine the reconstruction efficiency for the branching
fraction measurement, MC samples of $B^+\to X(3872)(\to D^{*0}\bar
D^0)K^+$ and $B^0\to X(3872)(\to D^{*0}\bar D^0)K_S^0$ events are
generated for $D^{*0}\to D^0\gamma$ and for $D^{*0}\to D^0\pi^0$.  The
$X(3872)$ is generated with a mass of $3872.5{\rm~MeV}/c^2$ and a
width of $4.0{\rm~MeV}/c^2$.  The total MC efficiencies multiplied by
subdecay branching fractions are presented in Table~\ref{tab:results}.

The branching fraction, assumed to be equal for charged and neutral
$B$ modes, is
\begin{eqnarray}
\label{eqn:branching}
  {\cal B}(B\to X(3872)K)\times{\cal B}(X(3872)\to D^{*0}\bar D^0) 
  \nonumber\\
  = (0.80\pm0.20\pm0.10)\times 10^{-4},
\end{eqnarray}
where ${\cal B}(X(3872)\to D^{*0}\bar D^0)$ stands for the sum of
branching fractions for $X(3872)\to D^{*0}\bar D^0$ and
$X(3872)\to\bar D^{*0}D^0$.

We obtain a mass of $(3872.9^{+0.6\,+0.4}_{-0.4\,-0.5}){\rm~MeV}/c^2$
and a width of $(3.9^{+2.8\,+0.2}_{-1.4\,-1.1}){\rm~MeV}/c^2$.  The
difference in mass between the peak and the $D^{*0}\bar D^0$ threshold
is
\begin{equation}
\label{eqn:massdiff}
  \delta M=M_X-m_{D^{*0}}-m_{D^0}=(1.1^{+0.6\,+0.1}_{-0.4\,-0.3}){\rm~MeV}/c^2.
\end{equation}

As a cross-check, different shapes are used as a signal function to
fit the data.  Using a non-relativistic Breit-Wigner function
truncated at the threshold, we obtain a mass of
$(3873.4^{+0.6}_{-0.9}){\rm~MeV}/c^2$, a width of
$(4.3^{+2.5}_{-1.4}){\rm~MeV}/c^2$ and a signal yield of
$39.6^{+9.3}_{-8.1}$ events with a statistical significance of
$8.0\sigma$.  Using a Flatt\'e-like parameterization~\cite{x-hanhart},
with $g=0.3$ and $f_{\rho}=0.007$, we obtain
$E_f=(-14.9\pm2.0){\rm~MeV}/c^2$, consistent with expected
$E_f=-11{\rm~MeV}/c^2$~\cite{x-hanhart}, and a signal of $65\pm12$
events with a statistical significance of $8.8\sigma$.  The data
statistics are not sufficient to distinguish between other fitting
functions suggested in the literature~\cite{braaten}.

The peak mass in this mode, $M_X$, like that in the BaBar
analysis~\cite{x-babar}, should not be directly compared to the mass
of the peak seen in $J/\psi\pi^+\pi^-$, or to the mass of the
$X(3872)$ state itself.  In this analysis, to improve
signal/background separation, $D^0\gamma$ and $D^0\pi^0$ combinations
are each subjected to a mass window selection, and then a
mass-constrained fit. This procedure returns masses above $D^{*0}\bar
D^0$ threshold by construction, so the distribution (and the mass and
width of the peak) should be interpreted accordingly.  Efforts in this
direction have already appeared in the
literature~\cite{braaten,kalashnikova}, addressing a preliminary
version of the results in this paper.  It should be noted that the
mass measurement in our earlier paper~\cite{x-gokhroo}, while lacking
the mass constraint, is also nontrivial to interpret, due to the role
of the $D^*$ width, and interference, in the decay
amplitudes~\cite{braaten,kalashnikova}.

For the $Y(3940)$ state, the simultaneous fit yields $7\pm21\pm4$
signal events and we set an upper limit of
\begin{equation} 
\label{eqn:br.Y3940}
  {\cal B}(B\to Y(3940)K)\times{\cal B}(Y(3940)\to D^{*0}\bar D^0)
  < 0.67 \times 10^{-4}
\end{equation}
at 90\% CL.  By averaging the branching fractions of
Refs~\cite{Y3940-belle}~and~\cite{Y3940-babar}, we obtain ${\cal
  B}(B\to Y(3940)K)\times{\cal B}(Y(3940)\to\omega
J/\psi)=(0.51\pm0.11)\times10^{-4}$; combining this with the upper
limit~(\ref{eqn:br.Y3940}) we get $\frac{{\cal B}(Y(3940)\to\omega
  J/\psi)}{{\cal B}(Y(3940)\to D^{*0}\bar D^0)}>0.71$ at 90\% CL, to
be compared with the 90\% CL limits from Ref.~\cite{x3940}, ${\cal
  B}(X(3940)\to\omega J/\psi)<0.26$ and ${\cal B}(X(3940)\to
D^{*0}\bar D^0)>0.45$, thus $\frac{{\cal B}(X(3940)\to\omega
  J/\psi)}{{\cal B}(X(3940)\to D^{*0}\bar D^0)}<0.58$ with more than
90\% CL, an incompatibility that suggests that the $X(3940)$ and the
$Y(3940)$ are different states.

In summary, we find a significant near-threshold enhancement in the
$D^{*0}\bar D^0$ invariant mass spectrum in $B\to D^{*0}\bar D^0K$
decays.  The significance of this enhancement including systematic
uncertainties is $6.4\sigma$; significant signals are seen in both
$D^{*0}\to D^0\gamma$ and $D^0\pi^0$ modes.  Using a relativistic
Breit-Wigner, we obtain a mass of
$(3872.9^{+0.6\,+0.4}_{-0.4\,-0.5}){\rm~MeV}/c^2$ and a width of
$(3.9^{+2.8\,+0.2}_{-1.4\,-1.1}){\rm~MeV}/c^2$.  The difference
between the fitted mass and the $D^{*0}\bar D^0$ threshold is
$(1.1^{+0.6\,+0.1}_{-0.4\,-0.3}){\rm~MeV}/c^2$.  Note that a $D^{*0}$
mass constraint has been used in this analysis; the fitted mass of the
$X(3872)$ peak is $2.3\sigma$ lower than the value obtained by
BaBar~\cite{x-babar}, where a similar constraint was used.  For the
$Y(3940)$ state, we set an upper limit on the ${\cal B}(B\to
Y(3940)K)\times{\cal B}(Y(3940)\to D^{*0}\bar D^0)$ branching fraction
which suggests that the $X(3940)$ and the $Y(3940)$ are different
states.

We thank the KEKB group for the excellent operation of the
accelerator, the KEK cryogenics group for the efficient
operation of the solenoid, and the KEK computer group and
the National Institute of Informatics for valuable computing
and SINET3 network support.  We acknowledge support from
the Ministry of Education, Culture, Sports, Science, and
Technology (MEXT) of Japan, the Japan Society for the 
Promotion of Science (JSPS), and the Tau-Lepton Physics 
Research Center of Nagoya University; 
the Australian Research Council and the Australian 
Department of Industry, Innovation, Science and Research;
the National Natural Science Foundation of China under
contract No.~10575109, 10775142, 10875115 and 10825524; 
the Department of Science and Technology of India; 
the BK21 and WCU program of the Ministry Education Science and
Technology, the CHEP SRC program and Basic Research program (grant No.
R01-2008-000-10477-0) of the Korea Science and Engineering Foundation,
Korea Research Foundation (KRF-2008-313-C00177),
and the Korea Institute of Science and Technology Information;
the Polish Ministry of Science and Higher Education;
the Ministry of Education and Science of the Russian
Federation and the Russian Federal Agency for Atomic Energy;
the Slovenian Research Agency;  the Swiss
National Science Foundation; the National Science Council
and the Ministry of Education of Taiwan; and the U.S.\
Department of Energy.
This work is supported by a Grant-in-Aid from MEXT for 
Science Research in a Priority Area (``New Development of 
Flavor Physics''), and from JSPS for Creative Scientific 
Research (``Evolution of Tau-lepton Physics'').


\begin{thebibliography}{10}
\bibitem{mXaverage}
Our own average using the most recent measurements from Belle, 
BaBar, CDF and D\O~\cite{x-karim,babar-08-007,CDF-note9454,x-D0}.
\end{thebibliography}
\end{document}